\documentclass[11pt,letterpaper]{article}
\usepackage{graphicx,epsfig,amsfonts,amssymb}
\usepackage{bm}
\usepackage{times}
\usepackage{color}
\usepackage{fixltx2e}

\begin{document}

\title{Hierarchical and Matrix Structures in a Large Organizational Email Network: Visualization and Modeling Approaches}


\author{Benjamin H. Sims
\and
Nikolai Sinitsyn
\and
Stephan J. Eidenbenz\\
Los Alamos National Laboratory\\
Los Alamos, New Mexico 87545
}
\date{}

\maketitle

\begin{abstract}
This paper presents findings from a study of the email network of a large scientific research organization, focusing on methods for visualizing and modeling organizational hierarchies within large, complex network datasets. In the first part of the paper, we find that visualization and interpretation of complex organizational network data is facilitated by integration of network data with information on formal organizational divisions and levels. By aggregating and visualizing email traffic between organizational units at various levels, we derive several insights into how large subdivisions of the organization interact with each other and with outside organizations. Our analysis shows that line and program management interactions in this organization systematically deviate from the idealized pattern of interaction prescribed by ``matrix management.'' In the second part of the paper, we propose a power law model for predicting degree distribution of organizational email traffic based on hierarchical relationships between managers and employees. This model considers the influence of global email announcements sent from managers to all employees under their supervision, and the role support staff play in generating email traffic, acting as agents for managers. We also analyze patterns in email traffic volume over the course of a work week. 
\end{abstract}

\section{Introduction}

In this paper, we present results of our analysis of a large organizational email dataset, comprising nearly complete email traffic records for Los Alamos National Laboratory (LANL) over a period of several months.\footnote{This document is an extended version of \cite{prev} as submitted to \emph{Lecture Notes in Social Networks} (Springer).} Very few organizational communication networks of this scale have been analyzed in the literature. Analyzing such large email datasets from complex organizations poses a number of challenges. First, considerable work is required to parse large quantities of raw data from network logs and convert it into a format suitable for network analysis and visualization. Second, a great deal of care is required to analyze and visualize network data in a way that makes sense of complex formal organizational structures - in our case, 456 organizational units that are connected through diverse organizational hierarchies and management chains. Finally, it can be difficult to sort out the effects of email traffic generated by mass announcements and communications along management chains from the more chaotic, less hierarchical traffic generated by everyday interactions among colleagues.

This paper addresses these complexities in two ways. First, we demonstrate methods for understanding large-scale structural relationships between organizational units by using carefully thought-out visualization strategies and basic graph statistics. Second, we propose a power law model for predicting the degree distribution of email traffic for nodes of large degree that engage in mass emails along hierarchical lines of communication. This likely characterizes a significant portion of email traffic from managers (and their agents) to employees under their supervision.

\section{Analysis of Organizational Structure}

While many analysts have examined ways of extracting structural features from corporate email exchange networks, they have typically focused at the level of email exchanges between individuals (albeit sometimes large numbers of individuals), bringing little or no information about formal organizational structures into their analysis \cite{people, spectral, micro}. Aggregating relationships based on formal organizational structures offers another important level of insight, which can be particularly useful for managers and analysts interested in interactions among business units, capabilities, or functions rather than individuals. Automatically collected email data has significant advantages for capturing interactions among organizational units: although email does not capture all relevant interactions, it provides comprehensive coverage across the entire organization without the overhead involved in large-scale survey-based studies. In order to locate individuals within organizational structures, we used organizational telephone directory data to associate email addresses with low-level organizational units, and information from organization charts to generate mappings of these units to higher-level ones.

\subsection{Structural relationships between elements of the organization}

Our analysis of structural relationships within LANL focuses on two broad, cross-cutting distinctions: program vs. line organizations, and technical research and development functions vs. operations functions (safety, physical plant, etc.)

\begin{figure}
\centering
\includegraphics[width=3in]{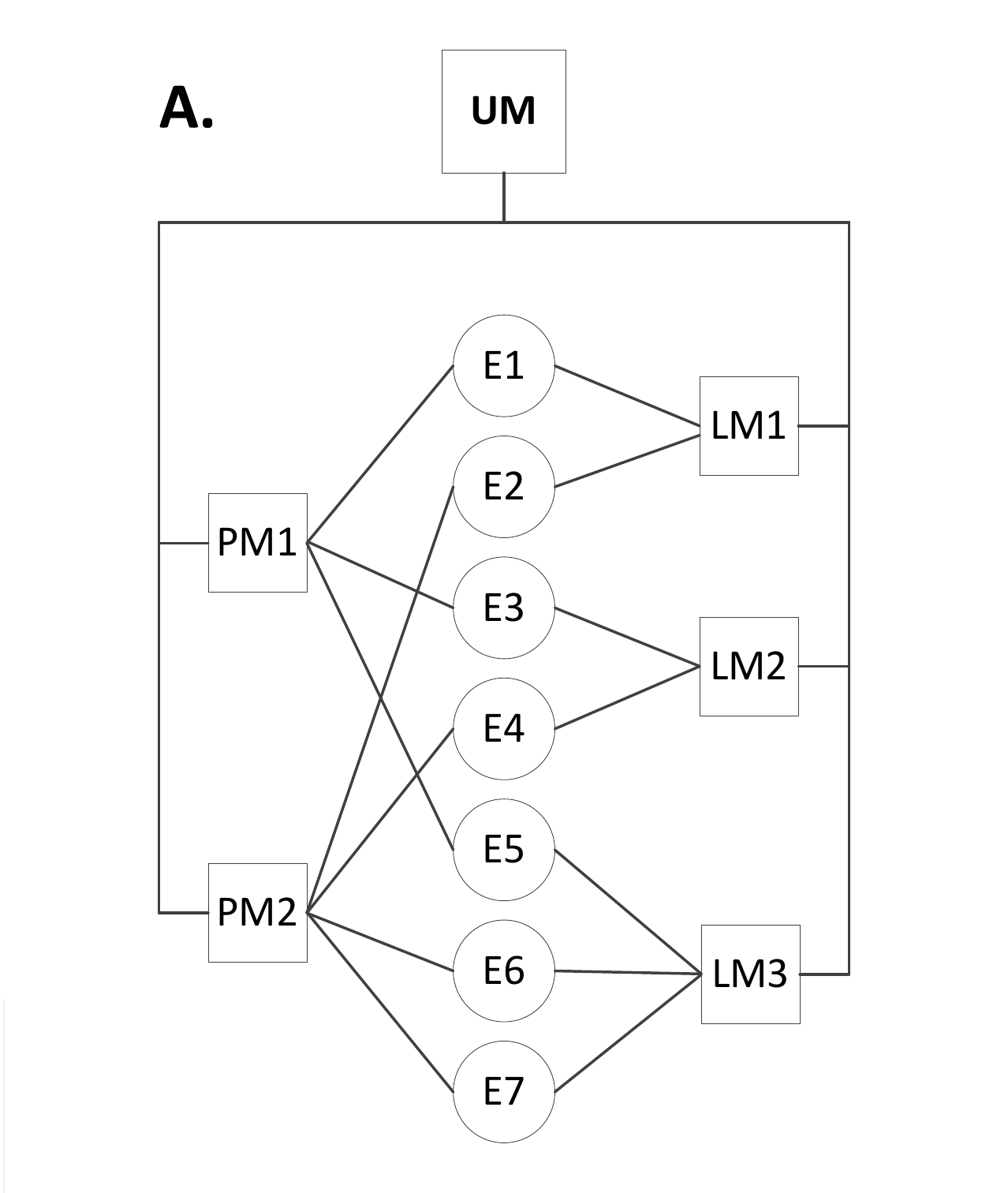}\\
\includegraphics[width=1.7in]{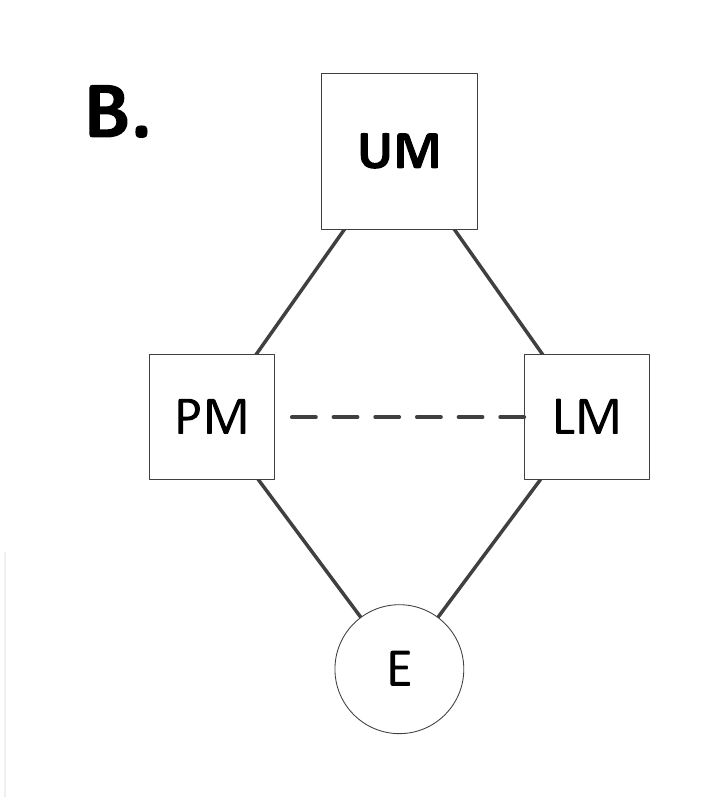}
\includegraphics[width=1.7in]{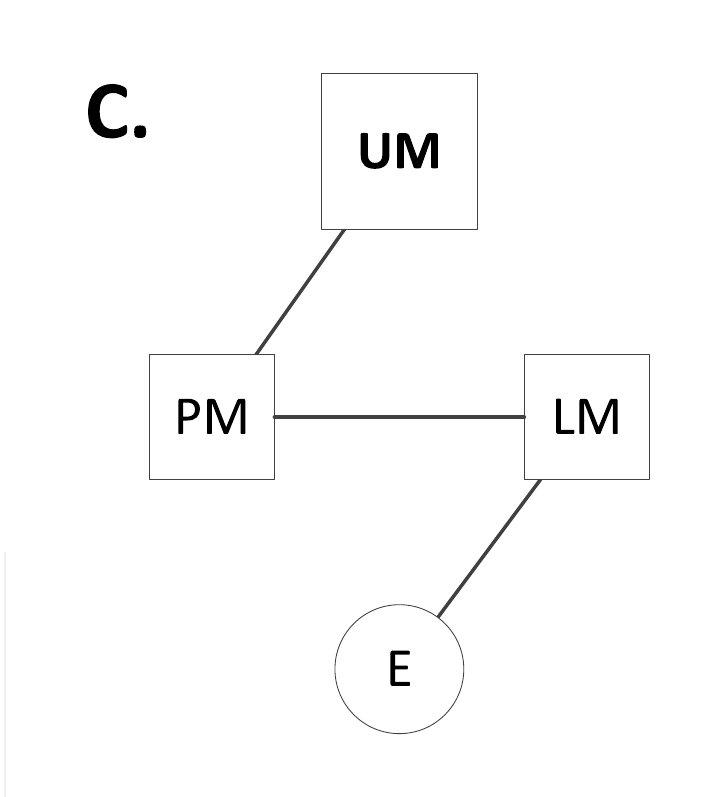}
\caption{A) Schematic representation of a typical organizational chart for a fully matrixed organization. Each employee reports to one line and one program manager, and line and program managers independently report to upper management. B) The idealized communication pattern that results from A. Dotted line indicates less frequent communication. C) The actual communication pattern at LANL, revealed through analysis of email data. (UM~= upper management, PM~= program/project management, LM~= line management, E~= employee.)}
\label{matrix}
\end{figure}

LANL is a hybrid matrix management organization. In a fully matrixed organization, each employee has two managers: a line manager and a program or project manager (Fig.~\ref{matrix}A). The employee is assigned to a line management unit based on their skill set and capabilities. For example, a computer scientist might be assigned to a Computational Modeling group, or an engineer to a Structural Engineering group. Line management plays little or no role in guiding the day-to-day work of employees, however. Instead, the employee is assigned to work on one or more projects, each of which is supervised by a program or project manager. A project is generally directed toward a specific product or deliverable, such design of a particular model of aircraft or completion of a particular research task. The day-to-day work of the employee toward these particular goals is directed by the program or project manager. Both line and program managers usually report, through some management chain, to upper level general managers. The idealized communication pattern that results is one in which program and line managers communicate primarily vertically, interacting with both upper management and employees (Fig.~\ref{matrix}B). In order to keep things running smoothly, however, program and line managers must also periodically communicate laterally, to ensure a good fit between capabilities and projects.

The matrix management model became popular in the aerospace industry with the rise of program management in the 1950s, and was in part influenced by the organizational structure of the Manhattan Project, \cite{bugos} in which Los Alamos played a major role. At LANL today, line and program organizations play less distinct roles. The base-level line units that house most employees are called groups, which may be built around programs or capabilities. In our analysis, we draw a distinction between groups and higher-level line management organizations, which aren't directly involved in technical or operations work. Program organizations play a variety of coordinating roles among groups, management, and outside organizations, and sometimes conduct technical or operations work as well. Despite this flexible definition, our analysis reveals that technical program organizations occupy a very well-defined structural space within the organization as a whole. 

\begin{figure}
\includegraphics[width=\textwidth]{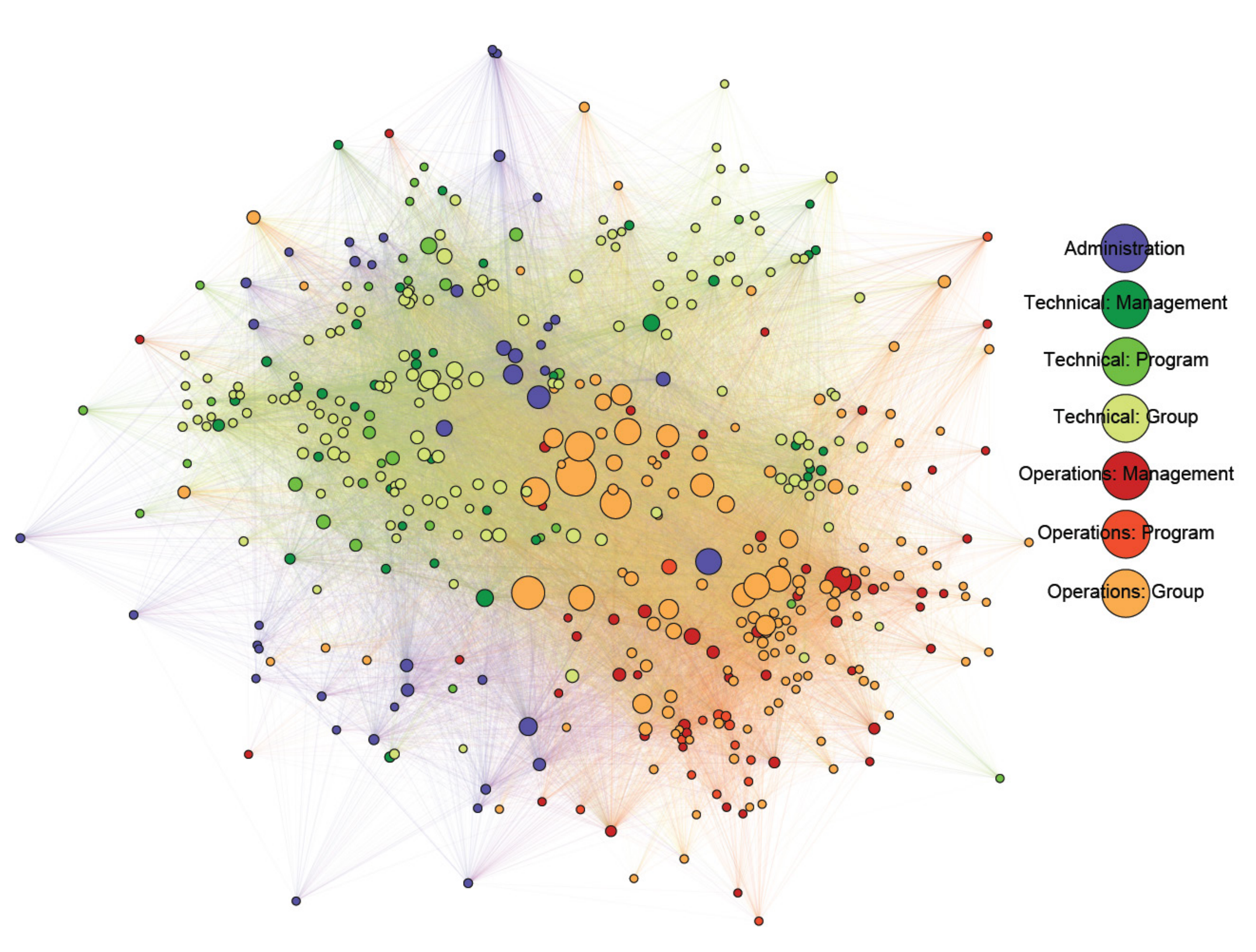}
 \caption{Email traffic between organizational units at LANL, using a force-vector layout. Node size represents betweenness centrality. Edge color is a mix of the colors of the connected nodes. Although individual edges are difficult to discern at this scale, the overall color field reflects the type of units that are most connected in a given region. }
 \label{biggraph}
 \end{figure}

Fig.~\ref{biggraph} shows email traffic between all organizational units at LANL over a period of 25 days, laid out using a force-vector algorithm. The units are colored according to the classification described above, and their sizes represent betweenness centrality. There are some visible patterns in this layout. First, a number of operations groups have the highest centrality, probably because they provide services to most of the other organizational units at the laboratory. Ranking the nodes by betweenness centrality confirms this: 17 of the top 20 nodes are operations organizations. In addition, operations units and technical units occupy distinct portions of the graph; this indicates that there is generally more interaction within these categories than between them. The highly central operations groups appear to play a bridging role between the two categories. Administration units appear to be somewhat more closely associated with technical units than operations units, although this is difficult to state with certainty.

\begin{figure}
\centering
\includegraphics[width=4in]{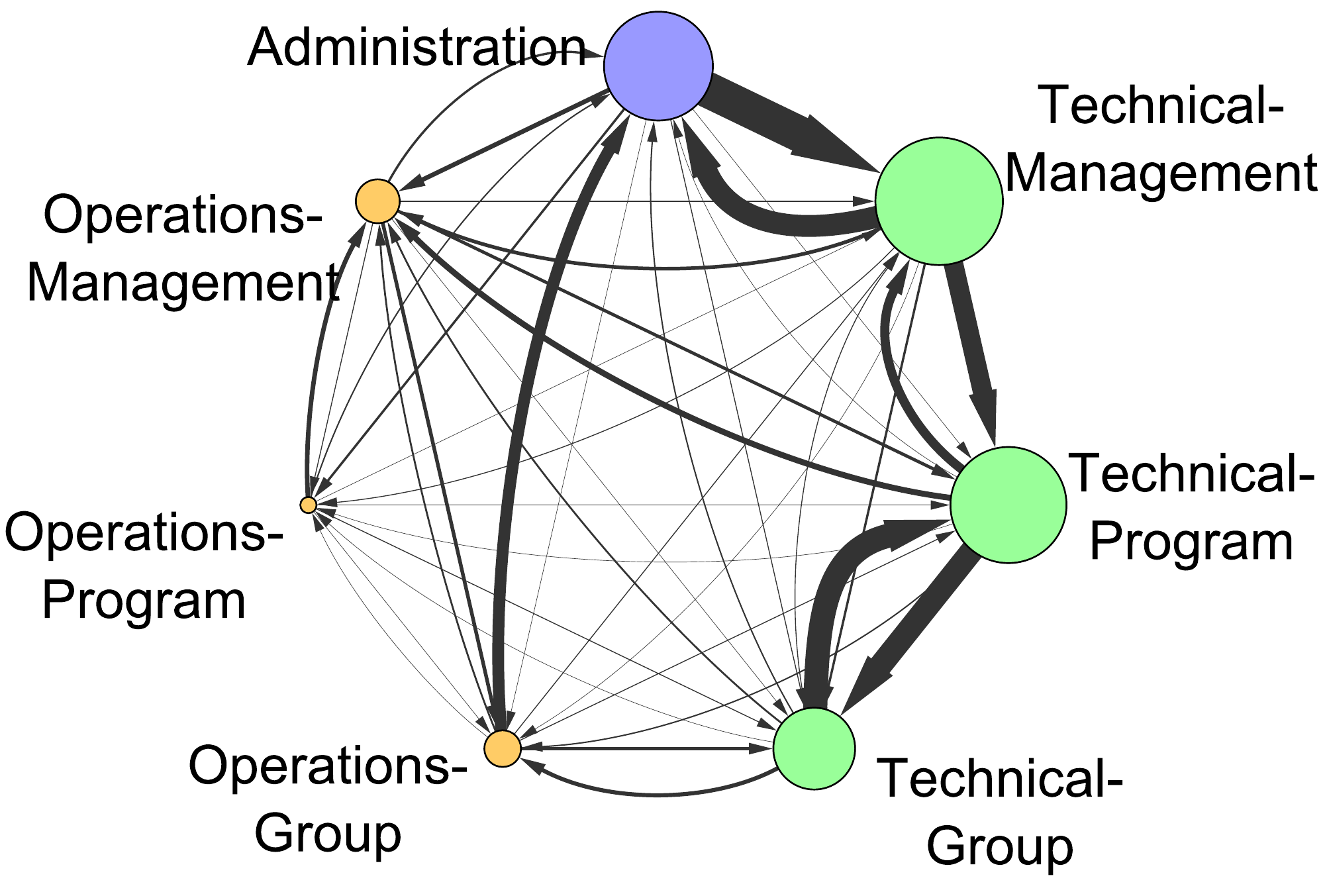}
\caption{Email traffic between organization types at LANL. Node diameter represents total degree (i.e. total number of incoming and outgoing emails) of the node; edge width represents email volume in the direction indicated. }
\label{topemail}
\end{figure}

Some of the ambiguities in interpretation can be clarified by grouping all units in a given category into a single node, resulting in the 7-node graph shown in Fig.~\ref{topemail}. This view, which uses a simple circular layout, reveals that there is a large amount of email traffic (in both directions) on the technical side of the organization along the path Administration - Management - Program - Group, and relatively little traffic between these entities along any other path. The operations side of the organization does not display this pattern, indicating that relationships between groups, programs, and management are more fluid there. This suggests that technical program organizations at LANL, rather than representing an independent chain of command (as in a true matrix organization) have instead evolved to play an intermediary role between technical groups and technical management. The structure of this relationship at LANL is depicted in Fig.~\ref{matrix}C. 

\begin{figure}
\centering
\includegraphics[width=4in]{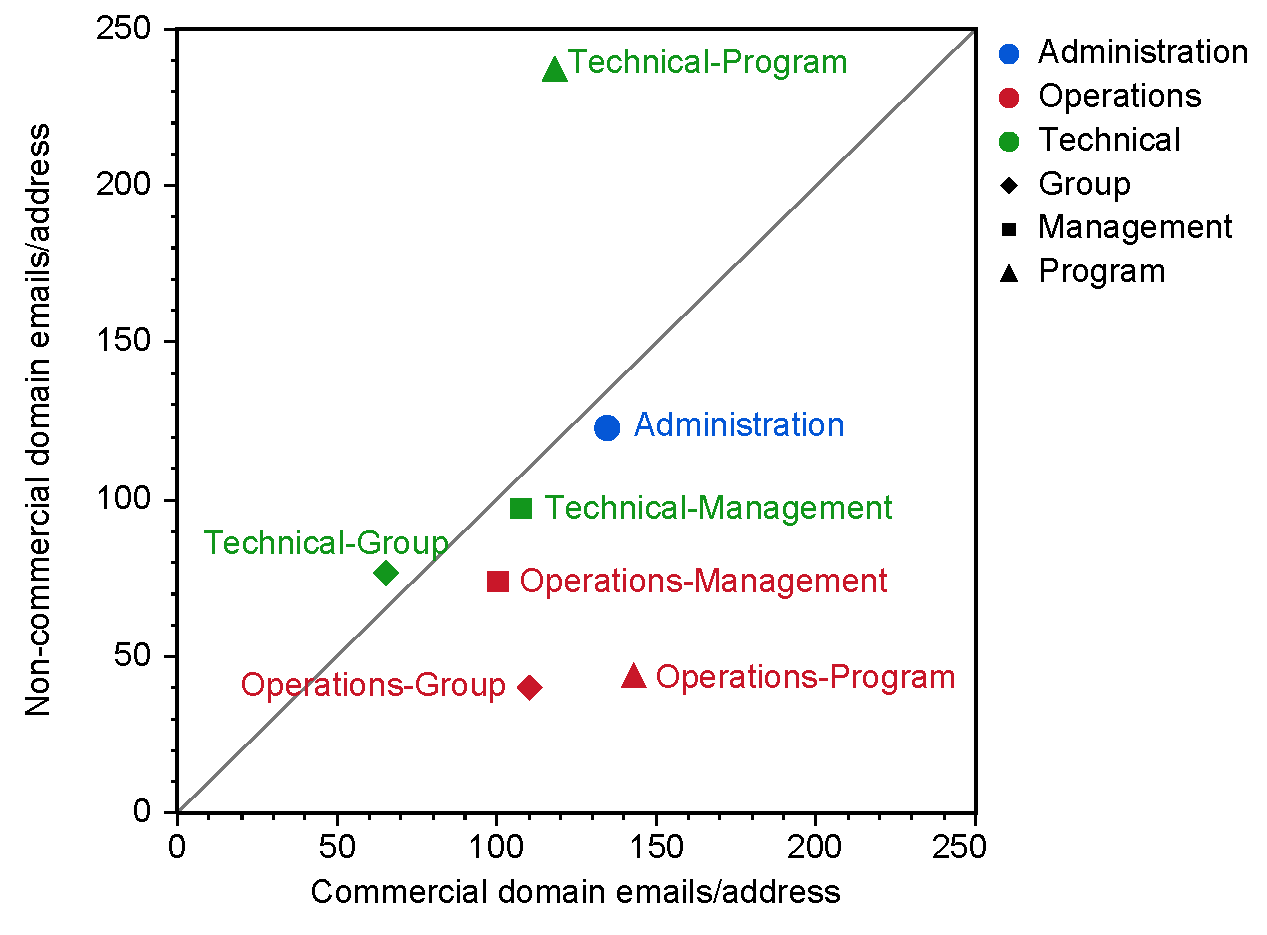}
 \caption{Total emails to/from commercial (.com, .net, .info) vs. non-commercial (.gov, .edu, .mil, etc.) domains, by organization type. } 
 \label{comnoncom}
 \end{figure}

Another way of understanding the roles different types of organizational units play is in terms of their relationships with outside entities. Fig.~\ref{comnoncom} plots the number of emails each type of organization sends and receives to/from commercial vs. non-commercial domains. This indicates that all types of operational units communicate significantly more with commercial entities, which is probably driven by relationships with suppliers and contractors. Technical groups, technical management, and administration communicate about equally with commercial and non-commercial domains. The outlier here is technical programs, which are much more highly connected to non-commercial domains, particularly .gov addresses. This further expands on the role of technical programs, suggesting that they are a nexus for coordination of technical work both internally, among line management organizations, and externally, between LANL and outside funding agencies. This is a potentially important finding, with implications for how program organizations should be supported and managed. 

\subsection{Structural relationships within organizational units}

\begin{figure}
\centering
\includegraphics[width=\textwidth]{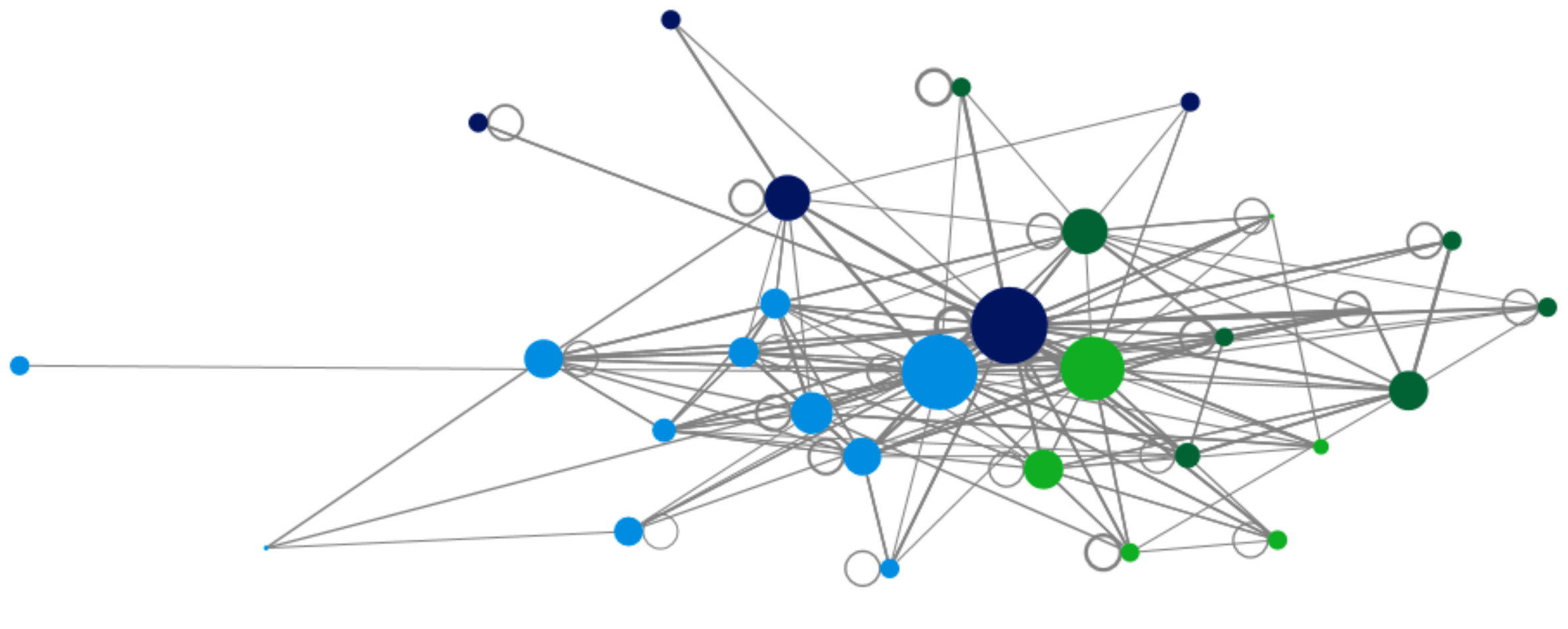}
 \caption{Email network for 2 week period in smaller group. Size of a node is proportional to logarithm of its betweenness centrality. Nodes with different colors correspond to different communities that were identified by application of the Girvan-Newman algorithm to the group's email network \cite{nodexl, girvan}. Link widths are proportional to the logarithm of the number of emails exchanged along these links. The network was visualized by assigning repulsion forces among nodes and spring constants proportional to the link weights, and then finding an equilibrium state.}
 \label{gsmall}
 \end{figure}

\begin{figure}
\centering
\includegraphics[width=\textwidth]{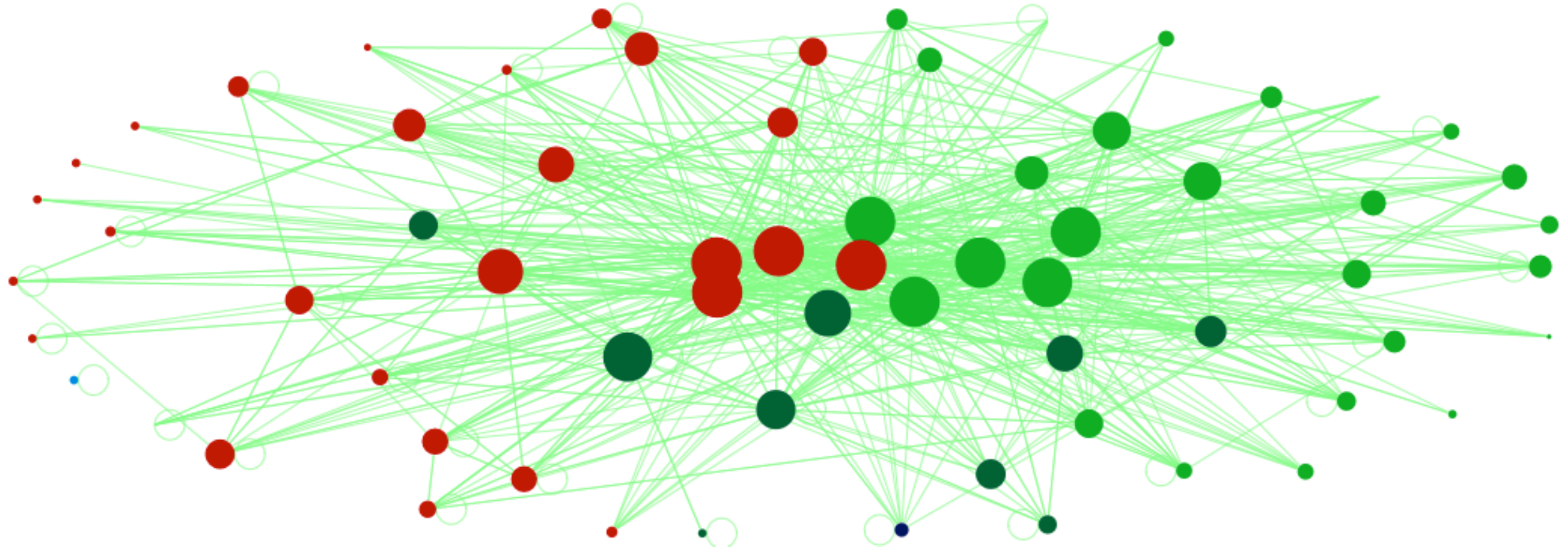}
 \caption{Email network  for 2 week period in larger group.}
 \label{gbig}
 \end{figure}

Email network maps can also be used to visualize relations among members of an organizational unit. Figures \ref{gsmall} and \ref{gbig} show email networks that were obtained from email exchange records among the members of two LANL groups over a period of two weeks. We intentionally chose groups that do similar work (theoretical research). In the smaller group in Fig.~\ref{gsmall}, the two nodes with highest betweenness centrality are group managers, and the third is technical support staff. Thus, the group has a relatively unified hierarchical structure with management and support staff at the center. In the larger group, managers were still among the most central nodes, but many other nodes had similar betweenness centrality (Fig~ \ref{gbig}). These include administrative assistants, seminar organizers, and several project leaders. This indicates a flatter, less centralized organizational structure. Application of a community detection algorithm to this graph reveals two main communities. As it happens, this group was created recently by merging two previously existing groups, and the detected communities generally correspond to those groups.


\section{Node connectivity distribution as a function of organizational hierarchy}

Several network types, including biological metabolic networks \cite{bar-bio}, the World Wide Web, and actor networks \cite{hier}, are conjectured to have power law distributions of node connectivity. In the case of metabolic networks, the interpretation of scale free behavior is  complicated by the lack of complete knowledge and relatively small sizes ($\sim$10$^3$ nodes) of such networks, while the mechanisms of self-similarity in many large social networks are still the subject of debate. However, organizational hierarchy has been shown to generate degree distributions for contacts between individuals that follow power laws \cite{bar}. 

Managers prefer to use email to communicate with subordinates in many different communication contexts \cite{markus}. We propose that node connectivity patterns in the email networks of large, formal organizations are driven, in part, by management hierarchy and specific patterns of email use by managers, in particular the mass broadcast of email announcements. Based on this observation, we develop a scale-free behavioral model that takes into account features specific to email communications in organizations. 
In this model, the self-similarity of the connectivity distribution of the email network is a consequence of the static self-similarity of the management structure, rather than resulting from a dynamic process, such as preferential attachment \cite{pr} or optimization strategies \cite{opt}. More specifically, self-similarity is due to the ability of a manager to continuously and directly communicate only with a relatively small number of people, while communications with other employees have to be conveyed in the form of broad announcements. 


Suppose that the top manager in an organization sends emails to all employees from time to time. This manager must correspond to the node in the email network that has highest connectivity $N$. 
Suppose that the top manager also talks directly (in person) to $l$ managers that are only one step lower in the director's hierarchy  (let's call them 1st level managers). Each of those 1st level managers, presumably, control their own subdivisions in the organization. Assuming roughly equal spans of managerial control, we can expect that, typically,  one 1st level manager sends emails to $N/l$ people. 
In reality, each manager also has a support team, such as assistants, administrators,  technicians, etc. who also may send announcements to the whole subdivision.


Let us introduce a coefficient $a$ which says how many support team employees are involved in sending global email announcements in the division on the same scale as their manager. We can then conclude that at the 1st level from the top there are $al$ persons who send emails to $N/l$ employees at a lower level. 

Each 1st level manager controls $l$ 2nd level ones and we can iterate our arguments, leading to the conclusion that there should be $(al)^2$ managers on the 2nd level who should be connected to $N/(l^2)$ people in their corresponding subdivisions. Continuing these arguments to the lower levels of the hierarchy, we find that, at a given level $x$, there should be $(al)^x$ managers (or their proxies) who write email announcements to $N/(l^x)$ people in their subdivision. 

Consider a plot that shows the number of nodes $n$ vs. the weight of those nodes, i.e. their outdegree $w$. Considering previous arguments, we find that the weight $w= N/( l^x)$ should correspond to $n=(al)^x$ nodes. Excluding the variable $x$, we find 
\begin{equation}
{\rm log}(n) = \frac{{\rm log}(al)}{{\rm log}(l)} \left({\rm log}(N) - {\rm log}(w) \right),
\label{lognw}
\end{equation}
where ${\log}$ is the natural logarithm.

Eq. (\ref{lognw}) shows that the distribution of connectivity, $n(w)$, in a hierarchical organizational email network should generally be a power law with exponent $\frac{{\rm log}(al)}{{\rm log}(l)}>1$. Obviously, at some level $x$, this hierarchy should terminate around the point at which $(al)^x= N/( l^x)$, because the number of managers should not normally exceed the number of employees.  Hence the power law (\ref{lognw}) is expected to hold only for nodes with heavy weights, e.g. $n>50$, i.e. for nodes that send announcement-like one-to-many communications, and at lower $n$ this model predicts a transition to some different pattern of degree distribution. At this level, it is likely that non-hierarchical communication patterns begin to dominate in any case.

\begin{figure}
\centering
\includegraphics[width=3in]{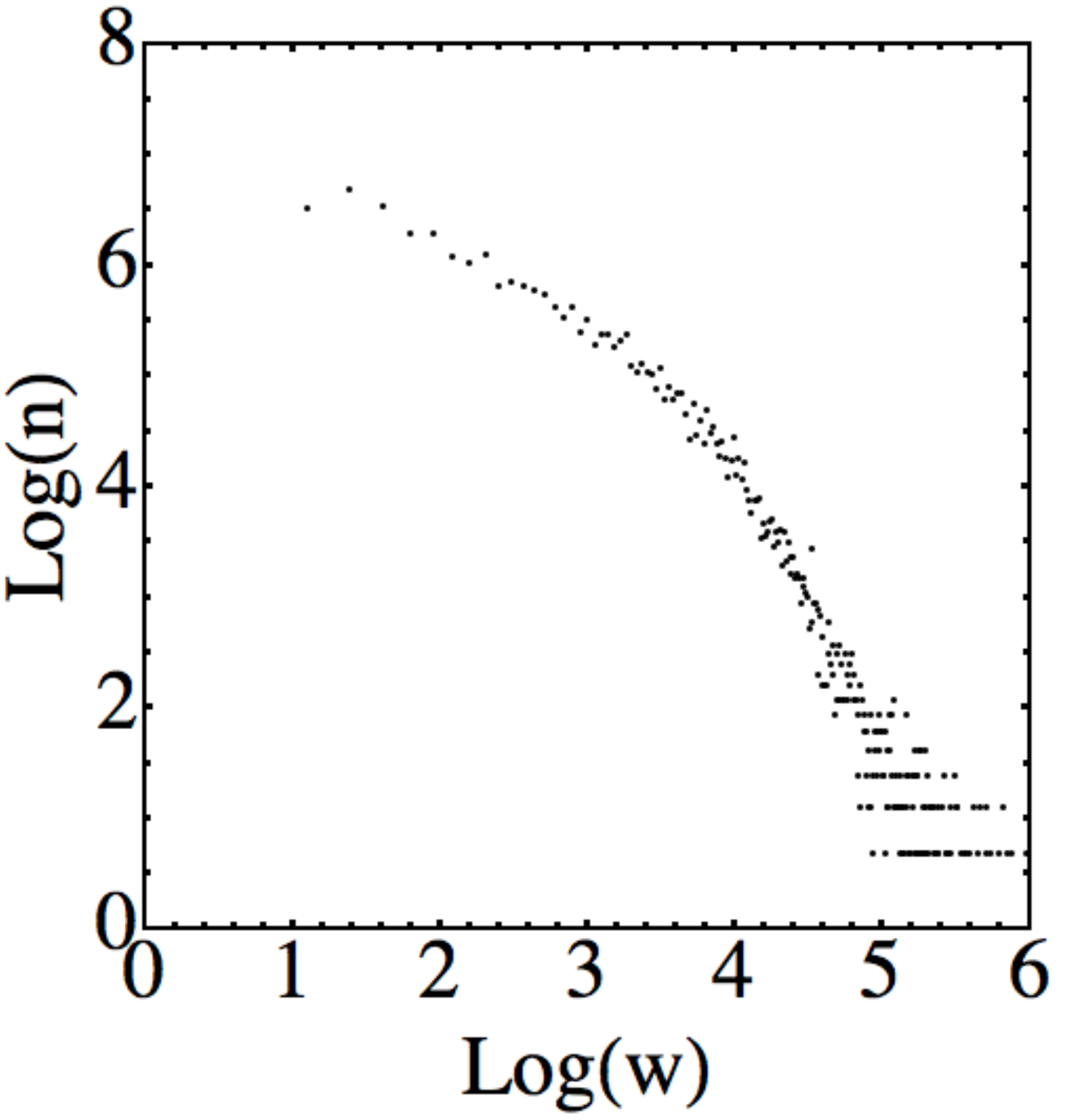}
 \caption{LogLog plot of the distribution of the number of nodes $n$ having the number of out-going links $w$. }
 \label{triv1}
\includegraphics[width=3in]{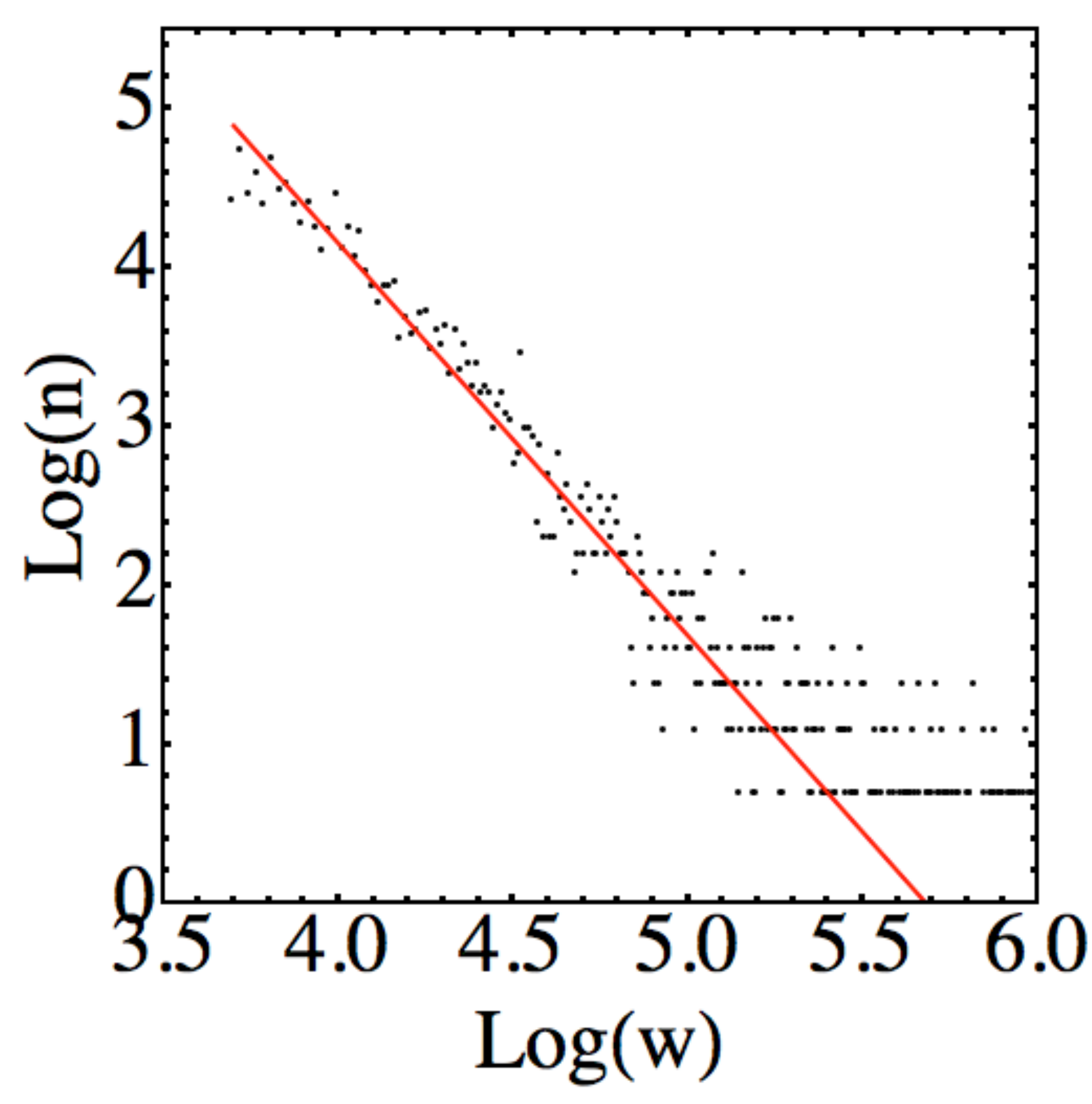}
 \caption{Zoom of Figure~\ref{triv1} for $w>40$. Red line is a linear fit corresponding to  ${\rm log}(n) \approx 14.0- 2.47 {\rm log}(w)$.}
 \label{triv2}
 \end{figure}

In order to compare this model to actual network data, we analyzed the statistics of node connectivity in email records at LANL during a two-week time interval (Fig.~\ref{triv1}). We removed nodes not in the domain $lanl.gov$ and cleaned the database of various automatically generated messages, such as bouncing emails that do not find their target domain. However, we kept domains that do not correspond to specific employees, such as emails sent from software support services. 
Our remaining network consisted of $N \approx 32000$ nodes, which is still about three times the number of employees at LANL.  This is partially attributed to the fact that we did not exclude domains that are not attached to specific people, and also the fact that a significant fraction of employees have more than one email address for various practical reasons.

Numerical analysis, in principle, should allow us to obtain information about parameters $l$, $x$ and $a$, from which one can make some very coarse-grained conclusions about the structure of the organization. Such an analysis should, of course, always be applied with a certain degree of skepticism due to potential issues with data quality, the simplicity of the model, and logarithmic dependence of the power law on some of these parameters \cite{power}.
We found that our data for $w>40$ could be well fitted by ${\rm log}(n) \approx 14.0- 2.47 {\rm log}(w)$ (Fig.~\ref{triv2}). If, e.g., we assume $l=4$,  then $a \approx 7$, i.e. each manager has the support of typically $a-1=6$ people, who help her post various announcements to her domain of control. The power law should terminate at the level of hierarchy $x$ given by $(al)^x= N/( l^x)$, which corresponds to $x \approx 3$, i.e. the email network data suggest that there are typically $x=3$ managers of different ranks between the working employee and the top manager of the organization. The typical number of email domains to which the lowest rank manager sends announcements is $w_{\rm min} \approx N/l^x \approx 48$. This should also  be the degree of the nodes at which the power law (\ref{lognw}) should be no longer justified. Indeed, we find the breakdown of the power law (\ref{lognw}) at $w<40$.   This estimate also predicts that a typical working employee receives emails from $(x+1)a=28$ managers or their support teams. 

Comparing these results to the actual organizational structure of the organization is very difficult due to the large excess of email addresses over the number of actual employees, and the lack of empirical data on many of the model parameters. Keeping in mind these difficulties, the estimated model parameters seem to be generally consistent with the actual organizational structure. In reality, LANL has 5 possible layers of line management between an employee and the laboratory director, but this is complicated by the facts that the lowest layer is often not used, and some employees work for organizations that report directly to a higher-level manager. So the estimate of $x \approx 3$ given above might be consistent with the actual organization structure. The average group size at LANL is difficult to determine quantitatively from available data, but appears to be generally in the 20-40 person range, which is somewhat lower than the number of domains (48) to which the lowest-level manager sends emails based on model estimates. Again, although these results might suggest possible conclusions about the accuracy of the model, we do not currently have data of sufficient quality to make a rigorous comparison between model estimates and real-world organizational structure in this case.

\section{Email traffic in real time}

\begin{figure}
\centering
\includegraphics[width=3in]{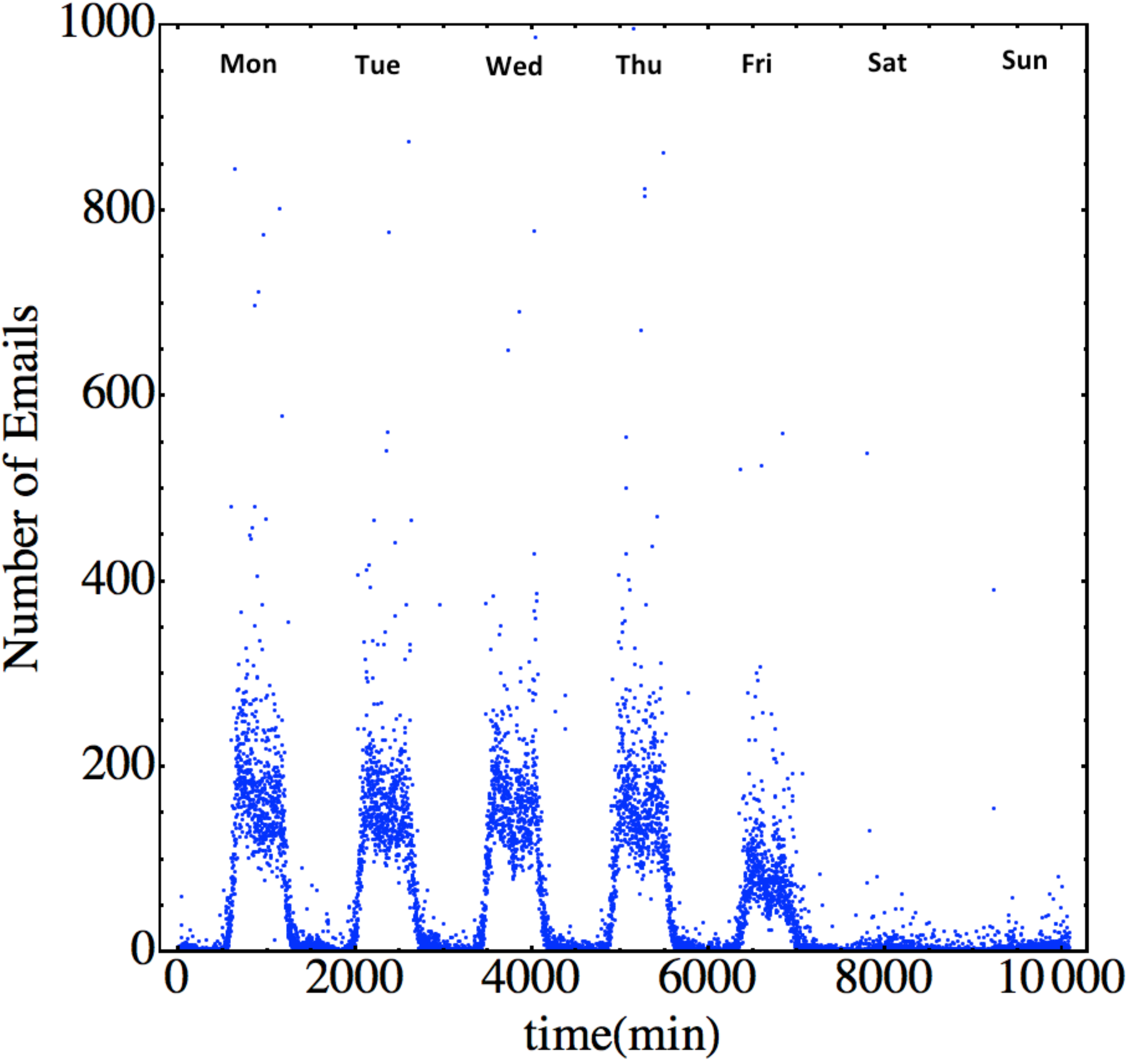} 
 \includegraphics[width=3in]{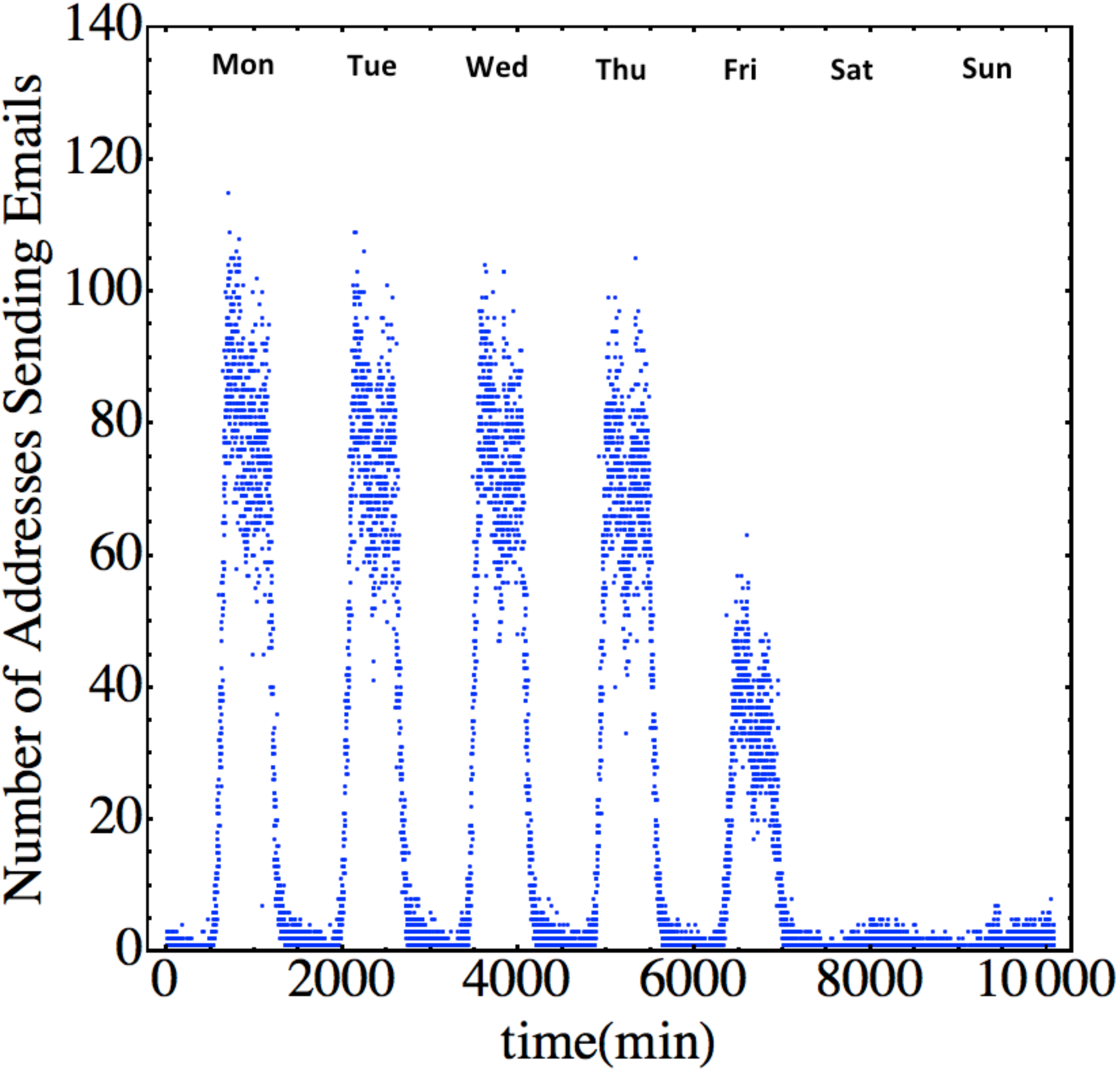}
 \caption{The number of emails sent per minute (top) and number of addresses sending email per minute (bottom) over a one week time interval.}
 \label{time1}
 \end{figure}
Fig.~\ref{time1} shows total email traffic and number of addresses sending email over one week with a one minute resolution. Working days have a bi-modal distribution with heaviest activity at the beginning and end of the day. The lower level of activity on Friday is related to an alternative work schedule that most LANL employees follow. This schedule enables employees to take every other Friday off in exchange for working longer hours Monday-Thursday. As a consequence, only slightly more than 50\% of the workforce is at work on a given Friday. This is directly reflected in the amount of email traffic on Fridays.

\section{Conclusion}

Visualizing and modeling email traffic in complex organizations remains a challenging problem. Visualizing email data in terms of formal organizational units reduces complexity and provides results that are more intelligible to organization members and analysts interested in understanding organizational structure at a macro level. For predicting the degree distribution of high-degree nodes in an organization, we find that it is useful to take into account both organizational hierarchy and email-specific behavior (in particular, the use of mass emails within line management chains). These findings suggest that considering  information about formal organizational structures alongside  email network data can provide significant new insights into the functioning of large, complex organizations.

\end{document}